\documentclass[12pt,letterpaper]{article}
\usepackage{sw20bams}

\input tcilatex

\QQQ{Language}{
American English
}

\begin{document}

\author{C. Viazminsky \\
IITAP, Iowa State University, Ames, IA 50011,USA\\
and Department of Physics, Aleppo University, Syria}
\title{Unitarily Equivalent Classes of First Order Differential Operators }
\date{June 27, 1999}
\maketitle

\begin{abstract}
The class of non-homogeneous operators which are based on the same vector
field, when viewed as acting on appropriate Hilbert spaces, are shown to be
isomorphic to each other. The method is based on expressing a first order
non-homogeneous differential operator as a product of a scalar function, a
differential operator, and the reciprocal scalar function.
\end{abstract}

\section{Introduction}

Let $M$ be an $n$-dimensional Riemannian manifold, $L$ be a continuous
vector field on $M$, and denote by $C^k(M)$ the set of real valued functions
which are differentiable of order $k$ on $M$. For each real valued
continuous function $q:M\rightarrow \Re $ there corresponds an operator $%
L+q:C^1(M)\rightarrow C^0(M)$ defined by $(L+q)\psi =L\psi +q\psi ,\;\forall
\psi \in C^1(M),$ where $L\psi $ is the Lie derivative of the function $\psi 
$ with respect to the field $L$, and $q\psi $ is the usual function
multiplication. It is evident that the operator $L+q$ from the vector space $%
C^1(M)$ to the vector space $C^0(M)$ is linear. Let $\eta $ be a real valued
function, differentiable on $M$, and such that $(L+q)\eta =0.$ We aim in
this work

(i) to show that the operators $L+q$ and $L$ are transformable to each other
by $L+q=\eta L\eta ^{-1}.$

(ii) to prove that, when defined on the appropriate Hilbert spaces of square
integrable functions on $M$, the operators $L+q$ and $L$ are isomorphic.

An extension of this work in the domain of totally linear partial
differential equation is given in \cite{Viaz}, and an algebraic treatment of
this extension is given in \cite{Aviaz}.

\section{Factorization of a First Order Non-Homogeneous Operator}

Let $\eta \in C^1(M)$ be a non-zero solution of the differential equation 
\begin{equation}
(L+q)\eta =0,  \label{e1}
\end{equation}
or equivalently, $\eta $ is any element in the kernel of the linear operator 
$L+q$ that is different from zero. We assume that $\eta $ has no zeros in $M$%
, and hence $\eta ^{-1}$ exists and of class $C^1(M)$. Now equation (\ref{e1}%
) is equivalent to 
\begin{equation}
q=-\eta ^{-1}(L\eta ).  \label{e2}
\end{equation}
We prove here a useful operator equality which will be used very often
through out this work.

\begin{proposition}
In $C^1(M)$ the following operator equality holds: 
\begin{equation}
\eta L\eta ^{-1}=L+q.  \label{e3}
\end{equation}
\TeXButton{Proof}{\proof}$\forall \psi \in C^1(M)$ we have 
\[
(\eta L\eta ^{-1})\psi =\eta (\eta ^{-1}L-\eta ^{-2}(L\eta ))\psi =(L-\eta
^{-1}(L\eta ))\psi =(L+q)\psi .
\]
We have used equation (\ref{e2}) when making the last step.
\end{proposition}

\begin{corollary}
Equality (\ref{e3}) is equivalent to 
\begin{equation}
L=\eta ^{-1}(L+q)\eta ,  \label{e4}
\end{equation}
which shows that all operators of the form $(L+q)$ which are constructed
from the same field $L$ may be transformed to $L$ and accordingly to each
other.
\end{corollary}

\begin{corollary}
From (\ref{e4}) we deduce that 
\begin{equation}
L^n=\eta ^{-1}(L+q)^n\eta \;\;\;and\;(L+q)^n=\eta L^n\eta ^{-1},\;
\label{e5}
\end{equation}
where $n$ is a non-negative integer. If $L$ is invertible then the latter
relation holds for all integers. We assume in relation (\ref{e5}) that $L$
is a $C^n$ vector field, $q$ and $\eta $ are $C^n$ and $C^{n+1}$ functions
respectively.
\end{corollary}

\begin{corollary}
If (\ref{e1}) holds, then it is easily checked that $(L+nq)\eta ^n=0,$ and
hence 
\begin{equation}
L=\eta ^{-n}(L+nq)\eta ^n.  \label{e6}
\end{equation}
In general, and for any real number $\alpha $, we have 
\[
L=\left| \eta \right| ^{-\alpha }(L+\alpha q)\left| \eta \right| ^\alpha . 
\]
\end{corollary}

\begin{corollary}
If $h$ is a real-valued continuous function on $M$ then $\eta
^{-1}(L+q+h)\eta =L+h$. It follows that 
\[
(L+h)\psi =0\Leftrightarrow (L+q+h)(\eta \psi )=0 
\]
\end{corollary}

\begin{corollary}
Take $h=-\lambda \;(\lambda \in \Re )$ in corollary (5) to obtain 
\[
(L-\lambda )\psi _\lambda =0\Leftrightarrow (L+q-\lambda )(\eta \psi
_\lambda )=0. 
\]
Expressed in words, the last relation states that: if $\psi _\lambda $ is an
eigenfunction of the operator $L$ belonging to the eigenvalue $\lambda $
then $\eta \psi _\lambda $ is an eigenfunction of the operator $L+q$
belonging to the same eigenvalue $\lambda .$

It is clear that the argument we have presented to obtain equality (\ref{e3}%
) can be carried through without any change to the case in which the set of
functions $C^k(M)$ is complex valued on $M$. The function $\eta $ is still
real so that the expression $\eta L\eta ^{-1}$ is real. From now on the set $%
C^k(M)$ will represent the set of complex valued functions which are
differentiable of order $k$ on $M$. It is obvious that theorem 1 and the
corollaries obtained from it are valid without modification.
\end{corollary}

\begin{example}
Take $L+q=\frac d{dx}+2x:C^1(\Re )\rightarrow C^0(\Re ).$ Since $\eta
=e^{-x^2}$ is a solution of (\ref{e1}), we have 
\[
\frac d{dx}+2x=e^{-x^2}\frac d{dx}e^{x^2}. 
\]
It is obvious that every complex number $\lambda $ is an eigenvalue of the
operator $\frac d{dx}$ to which an eigenfunction $\psi _\lambda =e^{\lambda
x}$ belongs. In accordance with the last corollary, it is easily checked
that $\lambda $ is an eigenvalue of the operator $\frac d{dx}+2x$ to which
the eigenfunction $\eta \psi _\lambda =e^{-x^2+\lambda x}$ belongs.
\end{example}

\section{Unitary Equivalence of L+ q and L.}

Let $(L^2,d\tau )\equiv H,$ where $d\tau $ is the volume element of $M$
defined by the Riemannian metric, be the natural Hilbert space of complex
valued square integrable functions on $M$. The inner product in $H$ is
denoted by $<,>$, and the associated norm by $\left\| {}\right\| ,$ so that
if $\psi \in H$ then $\left\| \psi \right\| ^2=\stackunder{M}{\int }\psi
^{*}\psi \;d\tau <\infty .$ Since the operator equality (\ref{e3}) is valid
on $C^1(M)$, it is also valid when restricted to the set of differentiable
functions in $H$ whose images under the action of either side of (\ref{e3})
is also in $H$. When restricted to $H$, the common domain of definition of
either side of (\ref{e3}), is 
\begin{equation}
D(L+q)=\{\psi \in H:\psi \in C^1(M),(L+q)\psi \in H\}.  \label{e7}
\end{equation}

Denote by $(L^2,\left| \eta \right| ^{2\alpha }d\tau )\equiv H_\eta $ the
set of complex valued functions on $M$ such that 
\[
\psi \in H_\eta \Leftrightarrow \left\| \psi \right\| _\eta ^2\equiv 
\stackunder{M}{\int }\psi ^{*}\psi \left| \eta \right| ^{2\alpha }d\tau
<\infty . 
\]
It is clear that $H_\eta $ is a Hilbert space, in which the inner product is
defined by 
\[
<\psi ,\phi >_\eta =\stackunder{M}{\int }\psi ^{*}\phi \left| \eta \right|
^{2\alpha }d\tau . 
\]
It is also clear that 
\begin{equation}
\psi \in H_\eta \Leftrightarrow \left| \eta \right| ^\alpha \psi \in H.
\label{e8}
\end{equation}

Consider the operator 
\[
\left| \eta \right| ^\alpha :H_\eta \rightarrow H,\;\;\;\psi \rightarrow
\left| \eta \right| ^\alpha \psi . 
\]
Since 
\[
<\psi \mid \phi >_\eta =\stackunder{M}{\int }\psi ^{*}\phi \left| \eta
\right| ^{2\alpha }d\tau =\stackunder{M}{\int }(\left| \eta \right| ^\alpha
\psi )^{*}(\left| \eta \right| ^\alpha \phi )d\tau =<\left| \eta \right|
^\alpha \psi ,\left| \eta \right| ^\alpha \phi >, 
\]
$\left| \eta \right| ^\alpha $ is an isometric operator. It is therefore
linear and has an inverse 
\[
\left| \eta \right| ^{-\alpha }:H\rightarrow H_\eta ,\;\;\psi \rightarrow
\left| \eta \right| ^{-\alpha }\psi , 
\]
which is also an isometry \cite{Akhiezer}.

\begin{proposition}
The operator $p=iL$ defined in $H_\eta $ on the domain 
\begin{equation}
D(p)=\{\psi \in H_\eta ,\psi \in C^1(M),p\psi \in H_\eta \}  \label{e9}
\end{equation}
is isomorphic to the operator $P=i\left| \eta \right| ^\alpha L\left| \eta
\right| ^{-\alpha \text{ }}$ defined in $H$ on the domain 
\begin{equation}
D(P)=\{\psi \in H,\left| \eta \right| ^{-\alpha }\psi \in C^1(M),P\psi \in
H\}.  \label{e10}
\end{equation}
\TeXButton{Proof}{\proof}According to the definition of unitarily equivalent
operators \cite{Akhiezer} it is sufficient for the problem we consider to
prove that $D(P)=\left| \eta \right| ^\alpha D(p)$. Indeed 
\begin{eqnarray*}
\left| \eta \right| ^\alpha D(p)=\{\left| \eta \right| ^\alpha \psi \in
H:\psi \in D(p)\} \\
\ =\{\left| \eta \right| ^\alpha \psi \in H:\psi \in H_\eta ,\psi \in
C^1(M),L\psi \in H_\eta \}\;by(9) \\
\ =\{\left| \eta \right| ^\alpha \psi \in H:\psi \in C^1(M),\left| \eta
\right| ^\alpha L\psi \in H\}\;\;\;by(8) \\
\ =\{\phi \in H,\left| \eta \right| ^{-\alpha }\phi \in C^1(M),\left| \eta
\right| ^\alpha L\left| \eta \right| ^{-\alpha }\phi \in H\} \\
\ =D(P).
\end{eqnarray*}
In the last step we set $\phi =\left| \eta \right| ^\alpha \psi ,$ and use
the fact that $\eta $ is in $C^1(M)$ and has no zeros in $M$, and hence $%
\phi $ is in $C^1(M)$ if and only if $\left| \eta \right| ^{-\alpha }\phi $
is in $C^1(M)$.

Proposition 2, in essence, asserts that the replacement of the operator $P$
by $p$ amounts to replacing the volume element $d\tau $ in $M$ by $\left|
\eta \right| ^{2\alpha }d\tau .$
\end{proposition}

\section{The Case of Non-Homogeneous Symmetric Operators}

It is well known \cite{Mackey,Wan} that the operator $\widehat{P}=-i(L+q),$%
with $q=\frac 12divL,$ defined on the domain 
\begin{equation}
D_0(\widehat{P})=\{\psi \in C_0^1(M):\psi ,\widehat{P}\psi \in L^2(M)\}
\label{e11}
\end{equation}
is symmetric. The symbol $C_0^1(M)$ refers to the set of continuously
differentiable functions with compact support in $M$. When the field $L$ is
complete the symmetric operator $\widehat{P}$ admits a self-adjoint
extension, which we denote again by $\widehat{P}$, and is defined in $%
L^2(M)\equiv (L^2,d\tau )$ on the domain 
\begin{equation}
D(\widehat{P})=\{\psi \in AC(M):\psi ,\widehat{P}\psi \in L^2(M)\},
\label{e12}
\end{equation}
where $AC(M)$ denotes the set of absolutely continuous functions on $M$. By
the spectral theory \cite{Fano}, there corresponds to the self-adjoint
operator $\widehat{P}$ an one-parameter group 
\begin{equation}
U_t=\exp (i\widehat{P}t)=\exp (i\eta \widehat{p}\eta ^{-1}t)  \label{e13}
\end{equation}
of unitary transformation of $(L^2,d\tau ).$ Now it is clear that the family
of unitary operators $\{V_t=\eta ^{-1}U_t\eta ,t\in \Re \},$ acting in the
Hilbert space $H_\eta \equiv (L^2,\eta ^2d\tau ),$ forms an one-parameter
group of transformations of this space, which is isomorphic to the group $%
U_t $ that acts in $H\equiv (L^2,d\tau ).$ The infinitesimal generator of
the group $V_t,$ denoted by $\widehat{p}_0,$ is given by 
\begin{equation}
\widehat{p}_0=\frac d{dt}V_t\mid _{t=0}=\eta ^{-1}\frac d{dt}U_t\mid
_{t=0}\eta =\eta ^{-1}\widehat{P}\eta =\widehat{p}.  \label{e14}
\end{equation}
It follows that the isomorphic operators $\widehat{p}$ and $\widehat{P}$
generate one-parameter groups of $(L^2,\eta ^2d\tau )$ and $(L^2,d\tau )$
respectively, which are isomorphic to each other.

Still, another way of reaching the latter result, concerning the operator $%
\widehat{p},$ is based on the following observation. Let $(x^1,...,x^n)$ be
a chart on $M$ in which the metric form is $ds^2=g_{ij}dx^idx^j$, and set $%
g=det(g_{ij}).$ The volume element of $M$ generated by the metric is $d\tau =%
\sqrt{g}dx^1...dx^n.$ The divergence of the vector field $L=\xi ^i\partial
/\partial x^i$ is given by \cite{Eisenhart,Lawden} $divL=g^{-1/2}(g^{1/2}\xi
^i)_{,i}$ , where comma denotes differentiation with respect to a
coordinate. It is apparent that defining the operator $\widehat{p}$ as
acting in the Hilbert space $H_\eta \equiv (L^2,\eta ^2d\tau )$ is
equivalent to define in $M$ a weighted volume element by $\eta ^2d\tau $. We
shall denote $M$ when endowed with this new volume element by $M_\eta .$ The
divergence of the vector field $L$ in $M_\eta $ will be denoted by $DivL$.
Now 
\begin{eqnarray*}
DivL &=&(\eta ^{-2}g^{-1/2})(\eta ^2g^{1/2}\xi _i)_{,i}=divL+2\eta ^{-1}\xi
_i\eta _{,i} \\
&=&divL+2\eta ^{-1}(L\eta )=divL+2(-\frac 12divL)=0.
\end{eqnarray*}
Thus the vector field $L$ in $M_\eta $ is incompressible \cite{Abraham}. Now
the operator $\widehat{p}=-iL$ defined on the domain 
\[
D_0(\widehat{p})=\{\psi \in C_0^1(M):\psi ,\widehat{p}\psi \in L^2(M_\eta
)\}, 
\]
where $L^2(M_\eta )\equiv (L^2,\eta ^2d\tau ),$ is symmetric. If $L$ is
complete then the operator $\widehat{p}$ defined on $D(\widehat{p})=\{\psi
\in AC(M):\psi ,\widehat{p}\psi \in L^2(M_\eta )\}$ is self-adjoint. The
one-parameter group $V_t$ of $L^2(M_\eta )$ generated by the self-adjoint
operator $\widehat{p}$ is given by $V_t=\exp (i\widehat{p}t).$ Expanding $%
V_t $ and $U_t$ and comparing the resulting expressions we find that $%
U_t=\eta V_t\eta ^{-1},$ which shows that these two groups are isomorphic.

Related works on incompressible fields and on generalized quantum momentum
observables are found in \cite{Iviaz} and \cite{Qviaz} respectively.

\begin{example}
Let our manifold $M$ be the semi-axis $(0,\infty ),$ and consider the
one-parameter group of transformation of $M$, $x\in M\rightarrow X=xe^t\in
M, $ where $t\in \Re .$ The infinitesimal generator of this group is the
complete vector field $L=x\frac d{dx}.$ Since $divL=1$, the operator $%
\widehat{P}=i(x\frac d{dx}+\frac 12)$ defined on the domain (\ref{e12}) is
self-adjoint. Noting that $\eta =x^{-1/2}$ is a solution of (\ref{e1}) we
have $\widehat{P}=ix^{1/2}\frac d{dx}x^{1/2}.$ There corresponds to $%
\widehat{P}$ an one-parameter group of transformation $U_t=\exp
(x^{1/2}\frac d{dx}x^{1/2}t)$ of $L^2(0,\infty )$. The vector field in $%
M_\eta $ is incompressible, for $Div(x\frac d{dx})=\eta ^{-2}(\eta
^2x)_{,x}=0.$ Therefore the operator $\widehat{p}$ defined in 
\[
\{\psi \in AC(0,\infty ):\psi ,x\frac{d\psi }{dx}\in L^2((0,\infty )_\eta
)\} 
\]
is self-adjoint. Comparing the expansions of the one-parameter groups $U_t$
and $V_t=\exp (ixd/dx),$ we verify easily that $U_t=\eta V_t\eta ^{-1}.$
\end{example}

\section{Acknowledgment}

The author thanks Prof. James P. Vary for the various facilities he made
available for me during my visit to IITAP, and thanks the University of
Aleppo in Syria for financial support.


\begin{thebibliography}{99}
\bibitem{Akhiezer}  Akhiezer N\ I and Glazmann I M, 1966, Theory of Linear
Operators in Hilbert Spaces. Frederick Ungar Publishing Co., New York.

\bibitem{Fano}  Fano G, 1967, Mathematical Methods of Quantum Mechanics.
McGraw-Hill Book Company, New York.

\bibitem{Mackey}  Mackey G W, 1963, Mathematical Foundations of Quantum
Mechanics. Reading.

\bibitem{Wan}  Wan K K and Viazminsky C P, Progress Theor. Phys. Vol. 58,
1977.

\bibitem{Eisenhart}  Eisenhart L, 1964, Riemannian Geometry. Princeton
University Press.

\bibitem{Abraham}  Abraham R and Marsden J, 1978, Foundation of Mechanics,
Reading, Mass., Benjamin/Cummings Pub. Co.

\bibitem{Lawden}  Lawden D F, 1975, An Introduction to Tensor Calculus and
Relativity. Chapman and Hall, New York.

\bibitem{Viaz}  Viazminsky C P, Comments on Lagrange Partial Differential
Equation. Los Alamos National Laboratory xxx.lanl.gov (FA/9906085), 1999.

\bibitem{Aviaz}  Viazminsky C P, An Algebraic Treatment of Totally Linear
Partial Differential Equations. To be published in Hadronic Journal of
Mathematics.

\bibitem{Iviaz}  Viazminsky C P, Incompressible Fields in Riemannian
Manifolds. xxx.lanl.gov (math-ph/9905025 v2 Jul 1999).

\bibitem{Qviaz}  Viazminsky C P, On Generalized Incompressible Momentum
Observables. To appear in Progress of Theoretical Physics.
\end{thebibliography}
\end{document}